# Fully Stabilized Mid-Infrared Frequency Comb for High-Precision Molecular Spectroscopy

*Markku Vainio[1,2*] and Juho Karhu[1]*

*Corresponding Author: E-mail: markku.vainio@helsinki.fi

[1]Department of Chemistry, University of Helsinki, P.O. Box 55, FI-00014, Finland
[2]VTT Technical Research Centre of Finland Ltd., Centre of Metrology MIKES, P.O. Box 1000, FI-02044 VTT, Finland

**Abstract:** A fully stabilized mid-infrared optical frequency comb spanning from 2.9 to 3.4 µm is described. The comb is based on half-harmonic generation in a femtosecond optical parametric oscillator, which transfers the high phase coherence of a fully stabilized near-infrared Er-doped fiber laser comb to the mid-infrared region. The method is simple, as no phase-locked loops or reference lasers are needed. Precise locking of optical frequencies of the mid-infrared comb to the pump comb is experimentally verified at sub-20 mHz level, which corresponds to a fractional statistical uncertainty of $2\times10^{-16}$ at the center frequency of the mid-infrared comb. The fully stabilized mid-infrared comb is an ideal tool for high-precision molecular spectroscopy, as well as for optical frequency metrology in the mid-infrared region, which is difficult to access with other stabilized frequency comb techniques.

## 1. Introduction

Extension of the laser frequency comb technology to the mid-infrared (MIR) spectral region has gained significant interest during the past few years owing to a large number of existing and potential applications in molecular spectroscopy, attosecond science, and optical metrology.[1-3] Many of these applications require or benefit from full stabilization of the frequency comb. The optical frequencies of the comb teeth can be written as $\nu_n = f_0 + nf_r$, where the mode number $n$ is a large integer. The two radio frequencies $f_0$ and $f_r$ are the carrier-envelope offset frequency and the repetition frequency, respectively. For a fully stabilized comb both $f_0$ and $f_r$ are stabilized, ideally by locking them to an accurate frequency reference. This is required particularly in optical frequency metrology [4] and Fourier synthesis [5], but also with the most sophisticated molecular spectroscopy methods, such as dual-comb spectroscopy [6] and high-precision frequency comb assisted spectroscopy.[7-9]

Mid-infrared frequency combs are typically generated by nonlinear optics, such as difference frequency generation (DFG) or optical parametric oscillation. An efficient version of the DFG approach is to divide the spectrum of an amplified femtosecond near-infrared (NIR) frequency comb into two portions, which are used as the pump and signal beams for DFG.[10, 11] Since the pump and signal fields are derived from the same near-infrared frequency comb, the offset frequency of the generated mid-infrared (idler) comb is inherently zero. Hence, full stabilization of a DFG comb only requires stabilization of the repetition frequency. A drawback of this method is that the offset frequency cannot be easily tuned, which may complicate the use of the comb in, e.g., cavity-enhanced frequency comb spectroscopy.[12]

Synchronously-pumped optical parametric oscillators (SP-OPO) provide another efficient means to transfer femtosecond NIR frequency combs to the mid-infrared region. Watt-level output power [13] and tunable center frequency [14] can be obtained by using a singly-resonant SP-OPO (**Fig. 1a**). Full stabilization of the singly-resonant SP-OPO is complicated and requires additional frequency conversion steps and/or stable reference lasers.[5, 13, 15] Doubly-resonant SP-OPO tuned to degeneracy provides a simpler approach, and allows for generation of a broad *instantaneous* spectrum with modest pump power. As an example, a frequency comb spanning from 2.6 to 7.5 µm at −20 dB level has been demonstrated.[16] In the degenerate doubly-resonant SP-OPO, which is also referred to as divide-by-2 subharmonic SP-OPO or half-harmonic generator, the signal and idler combs become indistinguishable and inherently phase locked to the pump comb (**Fig. 1b**).[17-19] In other words, full stabilization of the MIR comb can easily be achieved by using a standard, commercially available, fully stabilized NIR frequency comb as the pump source. The repetition frequency of the MIR comb is the same as that of the pump comb, as we experimentally verify with high accuracy in this article. Unlike with other SP-OPOs, the offset frequency is intrinsically locked to that of the pump laser. The degenerate SP-OPO has two alternative offsets that fulfil the energy conservation: the offset frequency of the MIR comb $f_{0,s} = f_{0,i}$ is either $f_{0,p}/2$ or $f_{0,p}/2 + f_r/2$.[17] Depending on the cavity length detuning of the SP-OPO cavity, the SP-OPO deterministically operates in one of these two offset frequency states that are shifted by half a repetition frequency.[19] This feature is useful for, e.g., absolute frequency spectroscopy in the mid-infrared region, as demonstrated in Section 4 of this article.

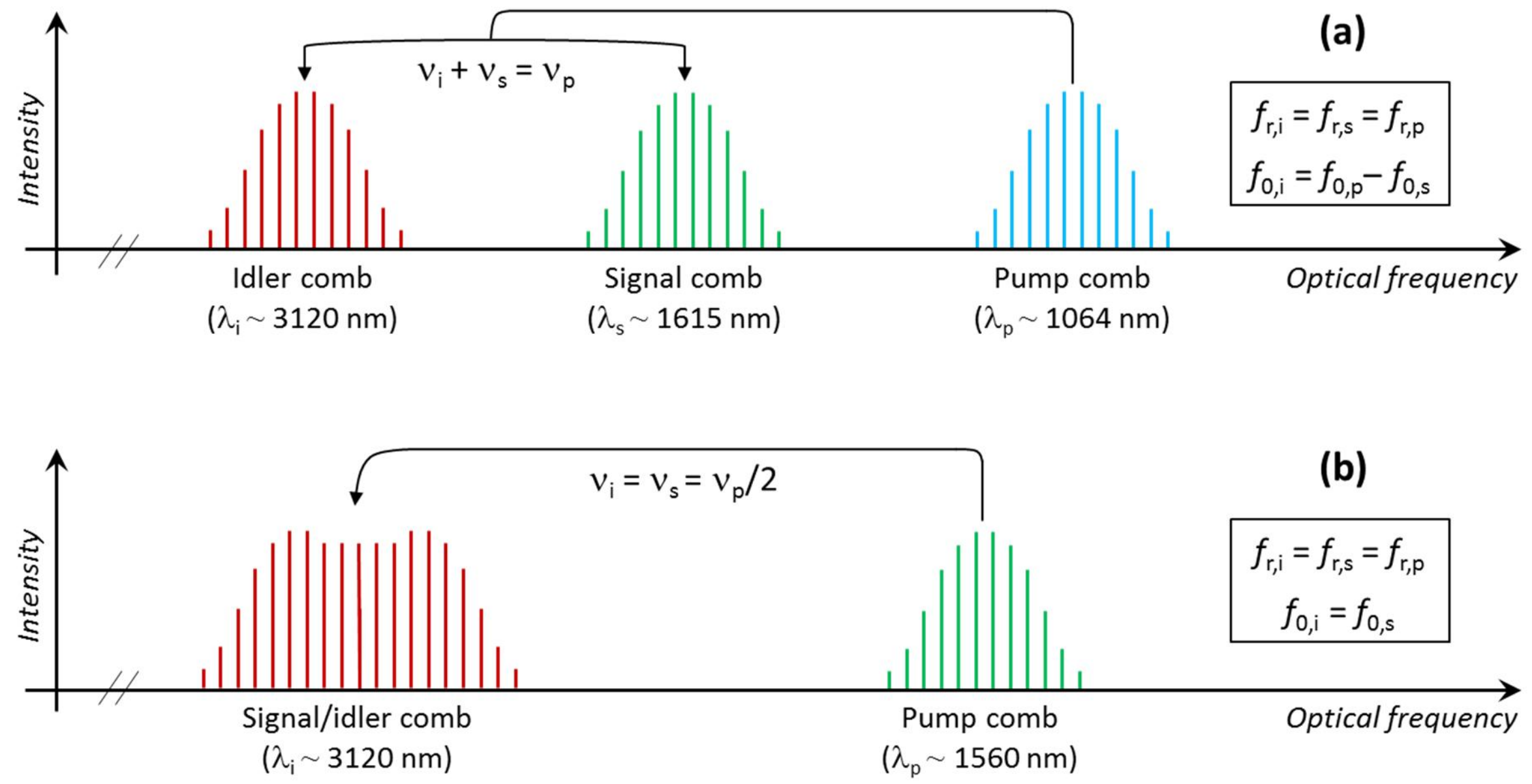

**Figure 1.** (a) Principle of mid-infrared (l $_i$ ~ 3 µm) frequency comb generation by singly-resonant OPO. The OPO is typically pumped with a mode-locked femtosecond laser (frequency comb source) at 1064 nm. The photon energy is conserved in the OPO process, such that $n_i = n_p - n_s$. (b) Mid-infrared comb generation by doubly-resonant degenerate OPO, which is pumped at 1560 nm. The pump frequency is halved (half-harmonic generation), such that $n_i = n_p/2$.

In the following section, we describe the first experimental demonstration of fully stabilized MIR frequency comb based on a degenerate SP-OPO pumped by an Er-doped fiber laser system. Rigorous characterization of the frequency precision of the fully stabilized MIR comb is presented in Section 3. An example of how the system can be applied to comb-assisted sub-Doppler spectroscopy of acetylene in the 3 µm wavelength region is given in Section 4.

## 2. Experimental Setup

The experimental setup for generation of a fully stabilized MIR frequency comb by a degenerate doubly-resonant SP-OPO is shown in **Fig. 2**. The OPO is synchronously pumped with a commercially available fully stabilized frequency comb that has a center wavelength of c.a. 1560 nm (Menlo Systems GmbH, FC1500-250-WG). The Menlo Systems comb generator is based on a mode-locked Er-doped fiber laser, followed by a polarization maintaining Er-doped fiber amplifier. The average output power after the amplifier is up to 400 mW. The comb repetition frequency is 250 MHz, which is locked to a Global Positioning System (GPS) disciplined crystal oscillator that has a relative stability better than $5\times10^{-12}$ in 1 s, and an accuracy better than $8\times10^{-12}$ in 1 s. The offset frequency is locked to the same time base using self-referencing based on *f*-2*f* interferometry.[20] Throughout the experiments reported in this article, the offset frequency was stabilized to 20 MHz.

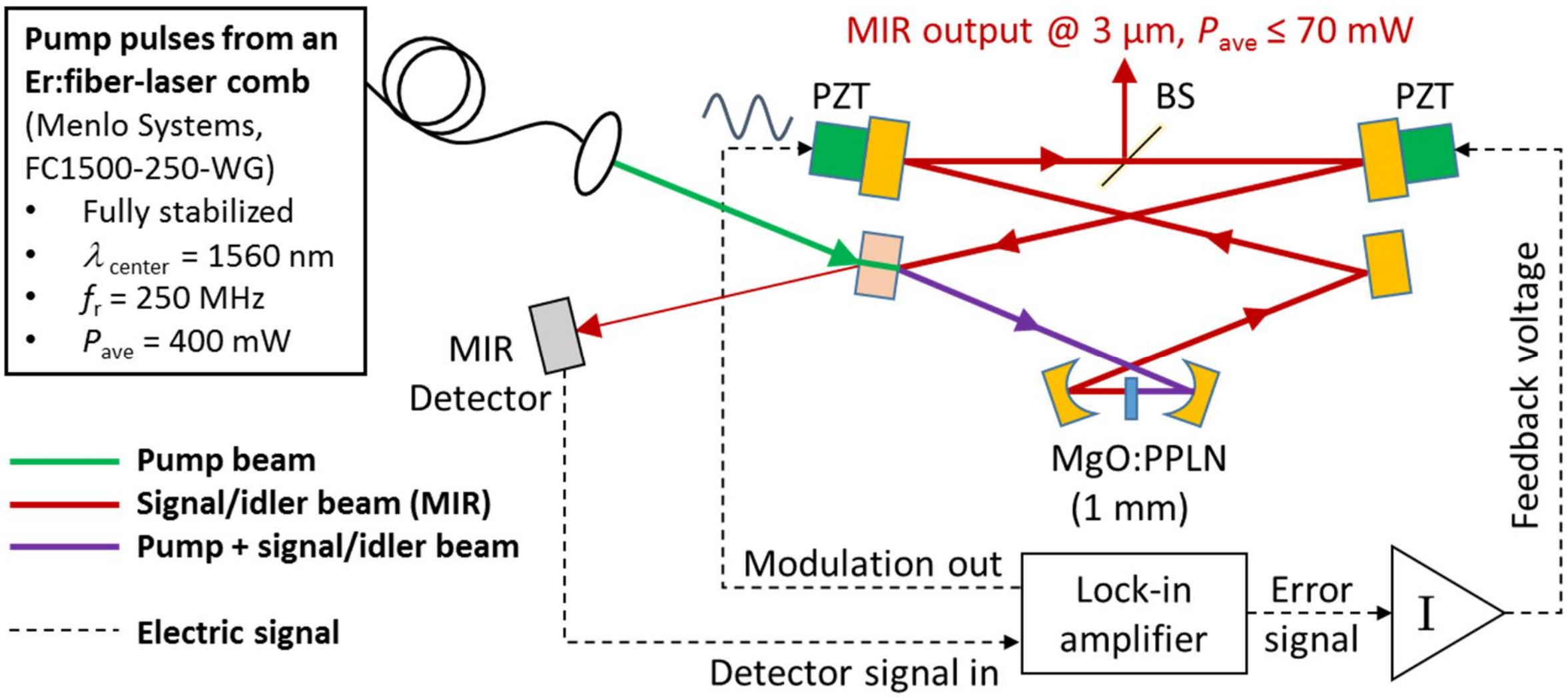

**Figure 2.** Schematic illustration of the degenerate SP-OPO used for mid-infrared comb generation. BS = beam splitter, PZT = piezoelectric actuator. No enclosure around the SP-OPO cavity was used during the experiments reported here.

The pump beam is coupled to free space and focused into a 1-mm long MgO-doped periodically poled lithium niobate crystal (MgO:PPLN, Covesion Ltd.), which is used as the nonlinear material of the SP-OPO. The crystal has a poling period of 35.2 µm, which is designed for type 0 phase-matching at room temperature. The crystal is antireflection coated for both the pump and mid-infrared comb wavelengths. The pump beam waist size ($1/e^2$-radius) in the crystal is 8 µm, which corresponds to a focusing parameter of c.a. 1.9. [21] The OPO resonator is designed to produce a mode-matched waist size of 11 µm in the crystal for the resonant mid-infrared beam. The resonator consists of 6 mirrors, all of which are gold coated except for the pump input mirror, which has a dielectric coating that is designed for high transmission of ~90% at the pump wavelength, and for high reflectivity of > 99% at the resonant mid-infrared wavelengths, between 2.9 and 3.4 µm (Laseroptik GmbH). Group delay dispersion (GDD) due to the mirror is between -200 and 300 $fs^2$ within this range. For comparison, the GDD of the MgO:PPLN is approximately -600 $fs^2$ at the center wavelength of the mid-infrared comb. No dispersion compensating components are used in the SP-OPO. The OPO resonator round-trip time is synchronized to the pump pulse repetition frequency, corresponding to a resonator round-trip length of $c/f_r$ ~ 1.2 m. The mid-infrared frequency comb is coupled out using a coated pellicle beam splitter (Thorlabs BP145B4), which is placed in a 45 degree angle, providing 30 to 40% output coupling for the resonant, P-polarized mid-infrared beam. As a result, up to 70 mW of average MIR output power is obtained with 370 mW average pump power coupled into the SP-OPO.

In order to meet the synchronous pumping condition of doubly-resonant femtosecond OPO, the resonator length has to be accurately controlled. Fine adjustment is done using a piezoelectric actuator (PZT), on which one of the flat resonator mirrors is mounted. Several oscillation peaks at discrete intervals are observed when scanning the resonator length around the optimum, as demonstrated by the inset of **Fig. 3**. Each oscillation peak corresponds to a different output spectrum, depending on the total intracavity dispersion.[22, 23] Figure 3 shows a typical example spectrum, which was measured with a high-resolution Fourier Transform Infrared spectrometer (FTIR, Bruker IFS 120 HR) when the SP-OPO was tuned to one of the oscillation peaks. The spectral coverage between 2.9 and 3.4 µm is limited by the reflection bandwidth of the dielectric cavity mirror.

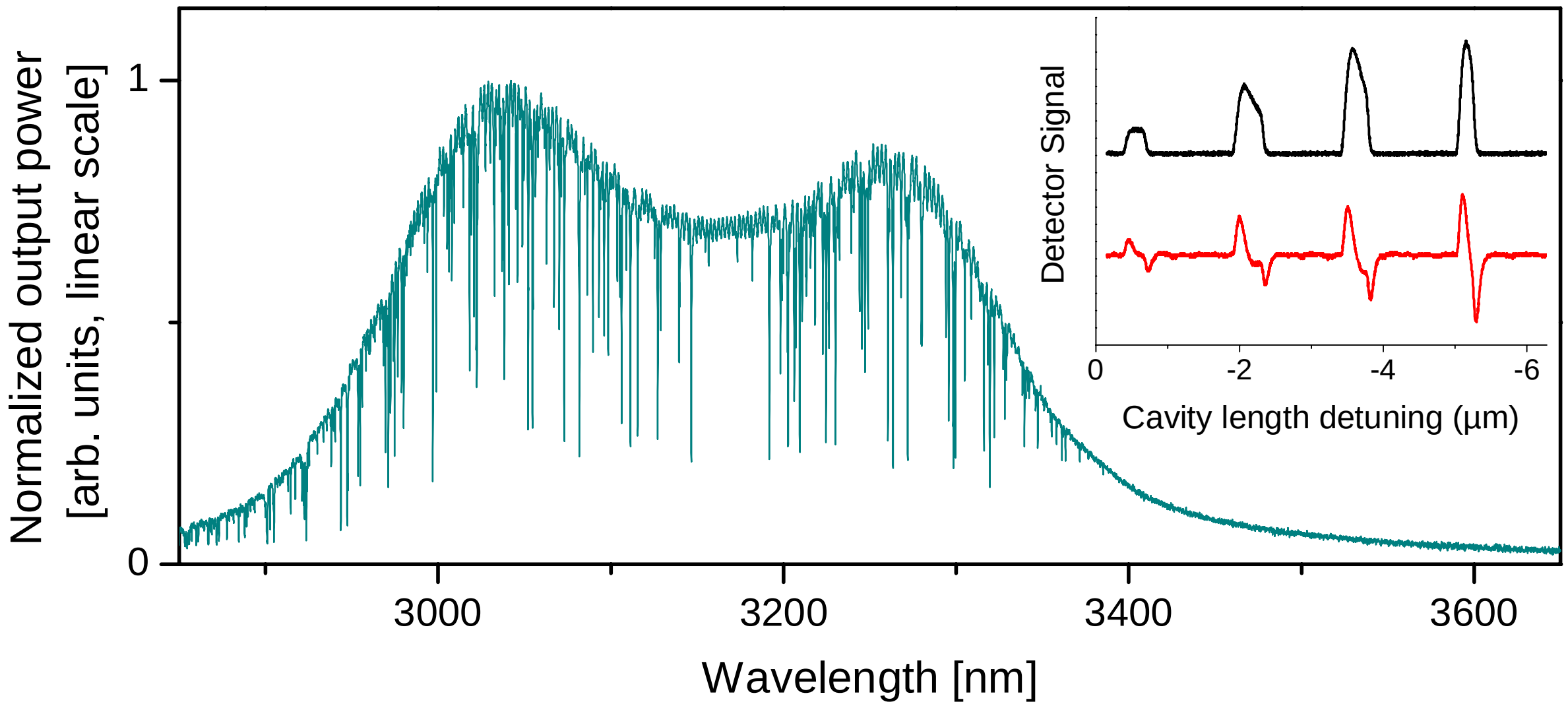

**Figure 3.** Envelope spectrum of the degenerate SP-OPO frequency comb. The absorption peaks are mostly due to water and methane in the air between the SP-OPO and the FTIR instrument that was used to record the spectrum. The inset shows oscillation peaks (in black) when tuning the SP-OPO cavity length. The associated 1*f*-signal used for cavity-length locking is shown by the red trace (see text for details).

Passive thermal locking owing to absorption of the resonant mid-infrared power in the MgO:PPLN crystal is sufficient to sustain SP-OPO oscillation for up to several minutes, but stable output power and reliable long-term locking over several hours or more requires the use of an active electronic locking scheme. Active locking is also important to maintain high spectral stability: The shape of the MIR output spectrum depends not only on the oscillation peak, but also on the locking point at each peak.[23, 24] In this work, we have applied the standard dither-and-lock method, see Fig. 2.[23, 25] A PZT actuator is used to produce a small (<10 nm) modulation of the resonator length. The modulation frequency is typically $f$ = 60-120 Hz, which allows us to use a compact and inexpensive thermopile (Heimann Sensor GmbH) for the detection of mid-infrared power that leaks out through the dielectric cavity mirror. The detected signal is demodulated with a lock-in amplifier, whose output gives the 1*f* error signal for a feedback circuit (integrator amplifier) that controls the SP-OPO resonator length via one of the PZT actuators. The integrator has a variable offset, such that the locking point can be adjusted, with zero offset corresponding to top-of-fringe locking. The locking bandwidth is set to below 10 Hz. We also tested larger locking bandwidths of up to c.a. 100 Hz, in which case a higher modulation frequency and a different MIR detector were used. This led to spectral instabilities, presumably due to interplay of the electronic lock with the thermal lock. Increase of the electronic locking bandwidth beyond that of the thermal lock, i.e. to the kHz level, could provide an alternative solution. However, in our experience, the slow feedback performs well and allows for stable output power over several hours. An example of output power stability over a measurement time of 1 hour and with a time resolution of 1 s is shown in **Fig. 4**. The output power was also recorded at a faster time scale using the Heimann thermopile sensor that has an electric bandwidth of c.a. 120 Hz. In this case, a standard deviation of the total average power between 0.7 and 2% was observed on the time scale of a minute, depending on the locking point.

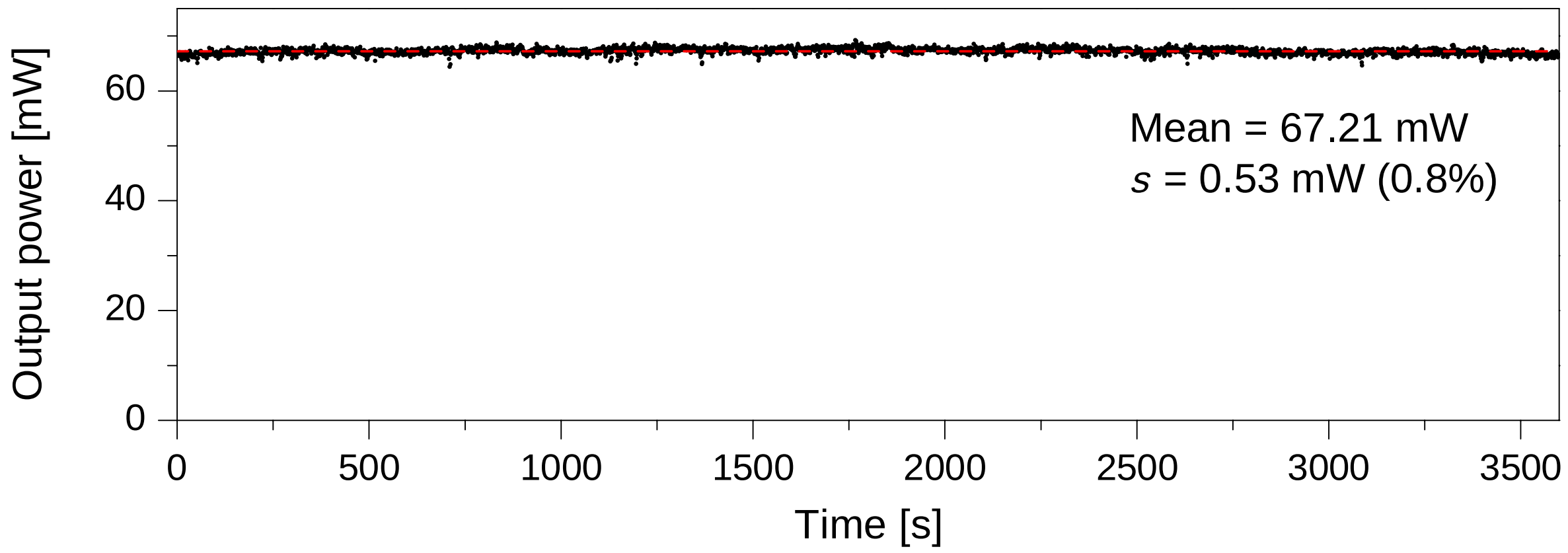

**Figure 4.** Stability of the average output power of the degenerate SP-OPO mid-infrared frequency comb over 1 h. The sampling interval is 1 s.

## 3. Frequency-domain characterization

The repetition frequency of the mid-infrared comb was measured by frequency doubling a portion of the mid-infrared spectrum, and subsequently counting the intermode beat frequency with a radio-frequency (RF) counter that was referenced to the same time base as the pump comb. The beat frequency was first mixed down to within the 50 MHz bandwidth of the zero-dead-time counter (K + K Messtechnik, model FXM50) that was used in the measurements. The accuracy of the repetition frequency was verified by comparing it with $f_r$ of the pump comb, which was recorded simultaneously. The results of this comparison for a measurement time of over 1 hour are summarized in **Fig. 5**. The measurement of the pump comb $f_r$ was essentially counter limited, the resolution of the counter being 0.25 mHz (1-s gate time). The measurement confirms precise reproduction of the comb repetition frequency in optical conversion from near-infrared to mid-infrared (and back to near-infrared) at 10-μHz level. This corresponds to a fractional statistical uncertainty of $4\times10^{-14}$ for a 1000-s averaging time.

Direct comparison of the frequency-doubled MIR comb with the NIR pump comb was carried out using the setup of **Fig. 6**. A continuous-wave (CW) external-cavity diode laser (ECDL, New Focus Velocity 6328) was first phase-locked to a tooth of the pump comb at 1552 nm. The offset frequency between the ECDL and the adjacent pump comb tooth was set to 21 MHz.

Beat frequency between the phase-locked ECDL and the frequency-doubled MIR comb was subsequently recorded. This comparison gives a precise measure of how well the mid-infrared comb is locked to the pump comb, since the noise of $f_r$ is magnified by twice the mid-infrared comb mode number $n$ ~ 385 000, and the possible instability of the offset frequency is also observed. An example of the beat note between the frequency-doubled MIR comb and the ECDL, as measured with an RF spectrum analyzer, is shown in the inset of Fig. 6. The observed linewidth of the beat note is 1 Hz, which is limited by the resolution bandwidth (RBW) of the spectrum analyzer. This implies sub-Hz linewidth of the mid-infrared comb teeth relative to the pump comb.

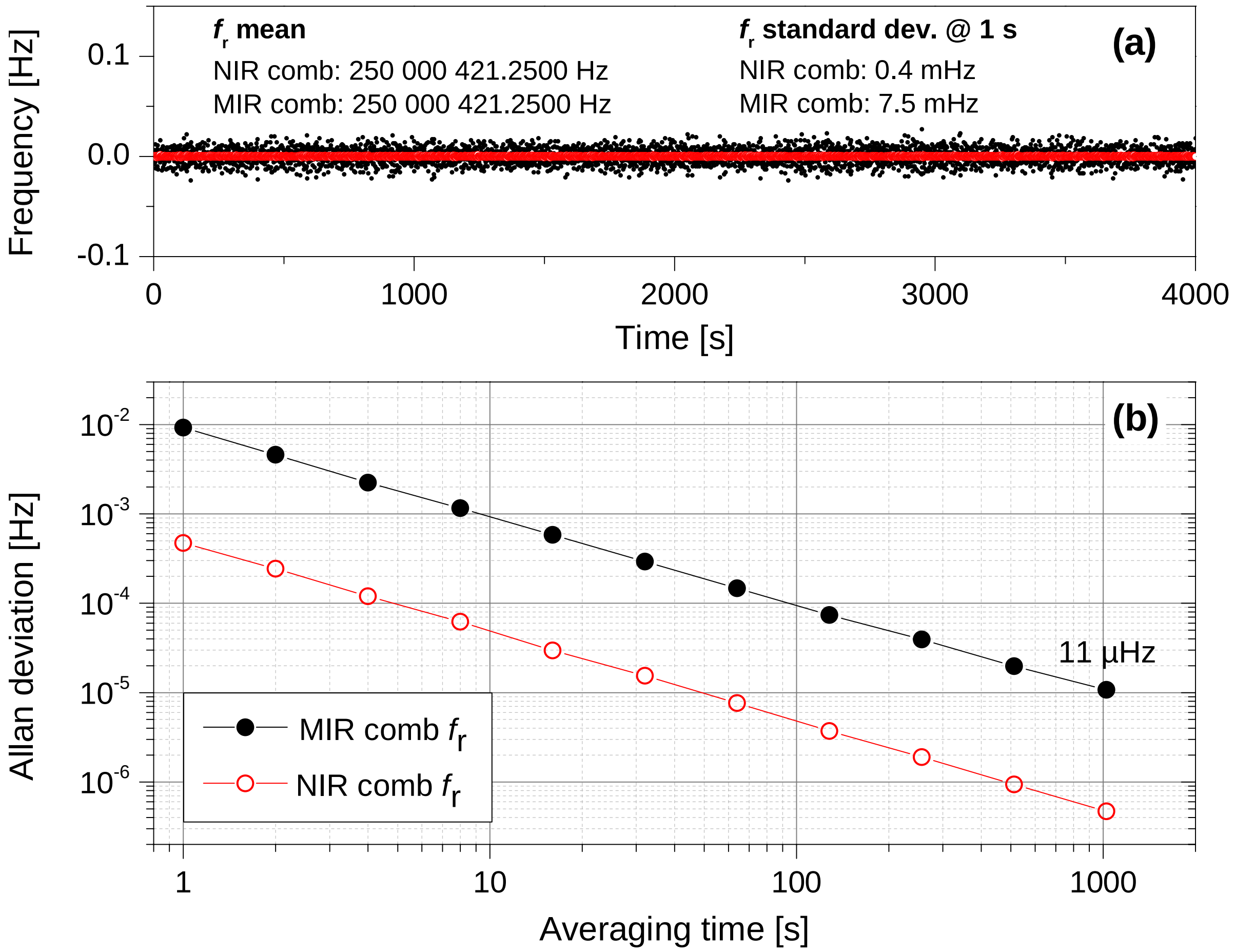

**Figure 5.** (a) Difference of the measured repetition frequency $f_r$ from the mean value as a function of measurement time. Solid black dots: Frequency-doubled MIR comb. Open red circles: NIR pump comb. A counter gate time of 1 s was used in the measurements. (b) The Allan deviation plots calculated from the time traces of panel (a).

The beat signal of the frequency-doubled MIR comb with the phase-locked ECDL was recorded with a frequency counter over a period of approximately one hour. A time trace of such measurement is shown in **Fig. 7a**, and the associated Allan standard deviation is plotted in **Fig. 7b**. For comparison, also shown are data from a similar measurement of the ECDL vs. the pump comb when the phase locking between them was established. The beat signals at 21 MHz were bandpass filtered (with two Minicircuits BBP-21.4+ filters) and amplified prior to counting. As can be seen from Fig. 7b, instability of the beat frequency between the frequency-doubled MIR comb and ECDL averages down to 34 mHz at 500 s. Again, it is worth noting that this includes the effect of the frequency-doubling step, which multiplies the frequency jitter of the mid-infrared comb teeth by a factor of two. Therefore, these measurements indicate that the teeth of the mid-infrared frequency comb are locked to the pump comb with a precision better than 20 mHz, which corresponds to a statistical uncertainty of $2\times10^{-16}$ at the center frequency $c$/3120 nm = 96 THz of the MIR comb. This means that the frequency accuracy and stability of the mid-infrared comb in applications such as dual-comb spectroscopy are not limited by the SP-OPO, but by the radiofrequency reference the pump comb is locked to, in agreement with the observations reported in Refs. [26, 27]. Note that this measurement also gives a more precise estimate for the instability of $f_r$ than the direct recording of $f_r$. Assuming uncorrelated $f_0$ and $f_r$, and a negligible $f_0$ instability, the upper limit for the repetition frequency instability of the MIR comb relative to the pump comb is 20 mHz/$n$ ~ 52 nHz (at 500 s).

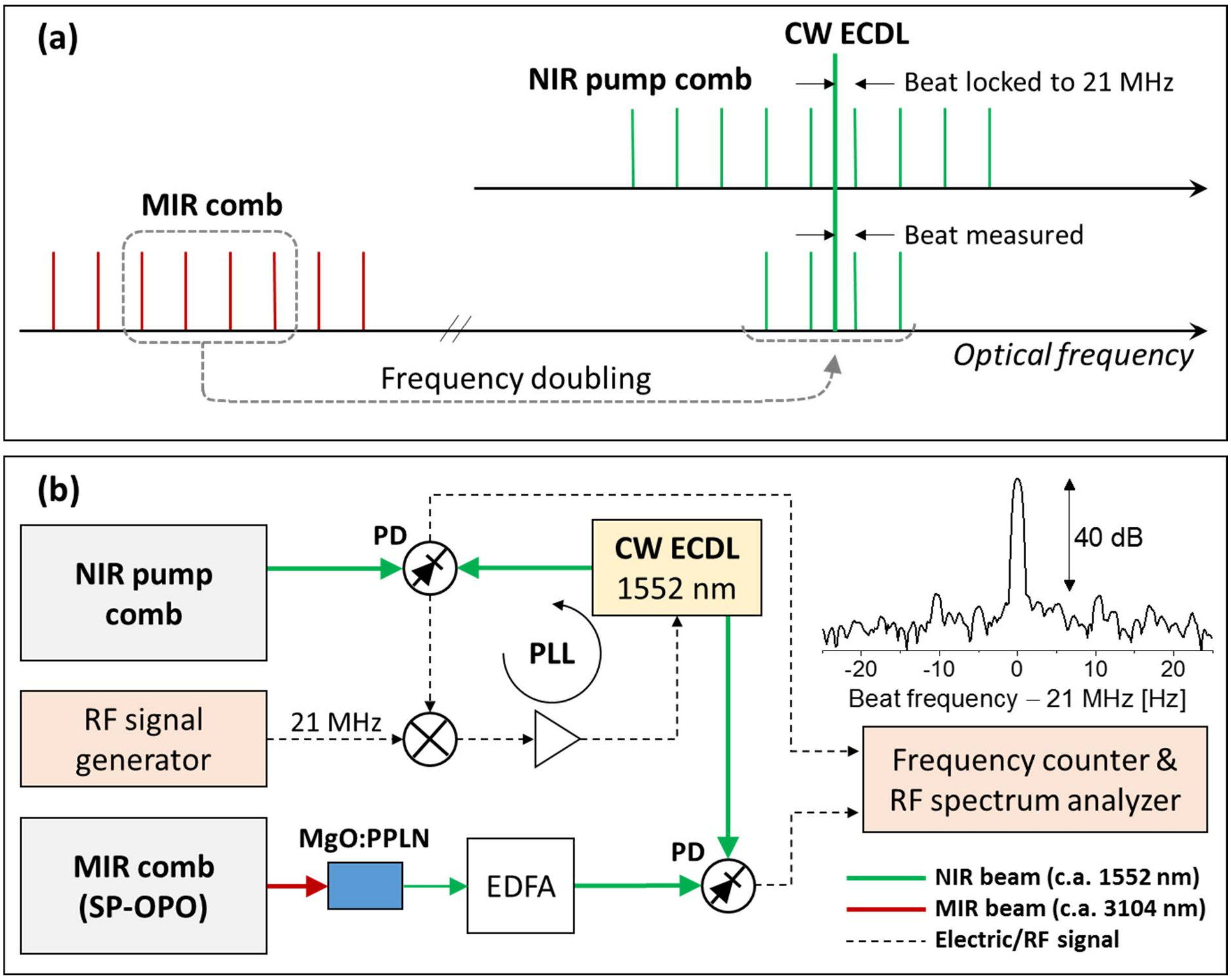

**Figure 6. (a)** Principle of frequency comparison of the MIR comb with the NIR pump comb. **(b)** Schematic of the setup used for the comparison. A portion of the MIR comb is frequency-doubled in an MgO:PPLN crystal to ~ 1552 nm. The frequency-doubled light is amplified in an Erbium:doped fiber amplifier (EDFA) and sent to a photodetector (PD) together with light from an external-cavity diode laser (ECDL), which is phase-locked to the NIR pump comb with a 21 MHz offset. The inset shows the resulting beat note measured with a 1 Hz resolution bandwidth. PLL = phase-locked loop. For simplicity, optical and RF filters etc. are not shown.

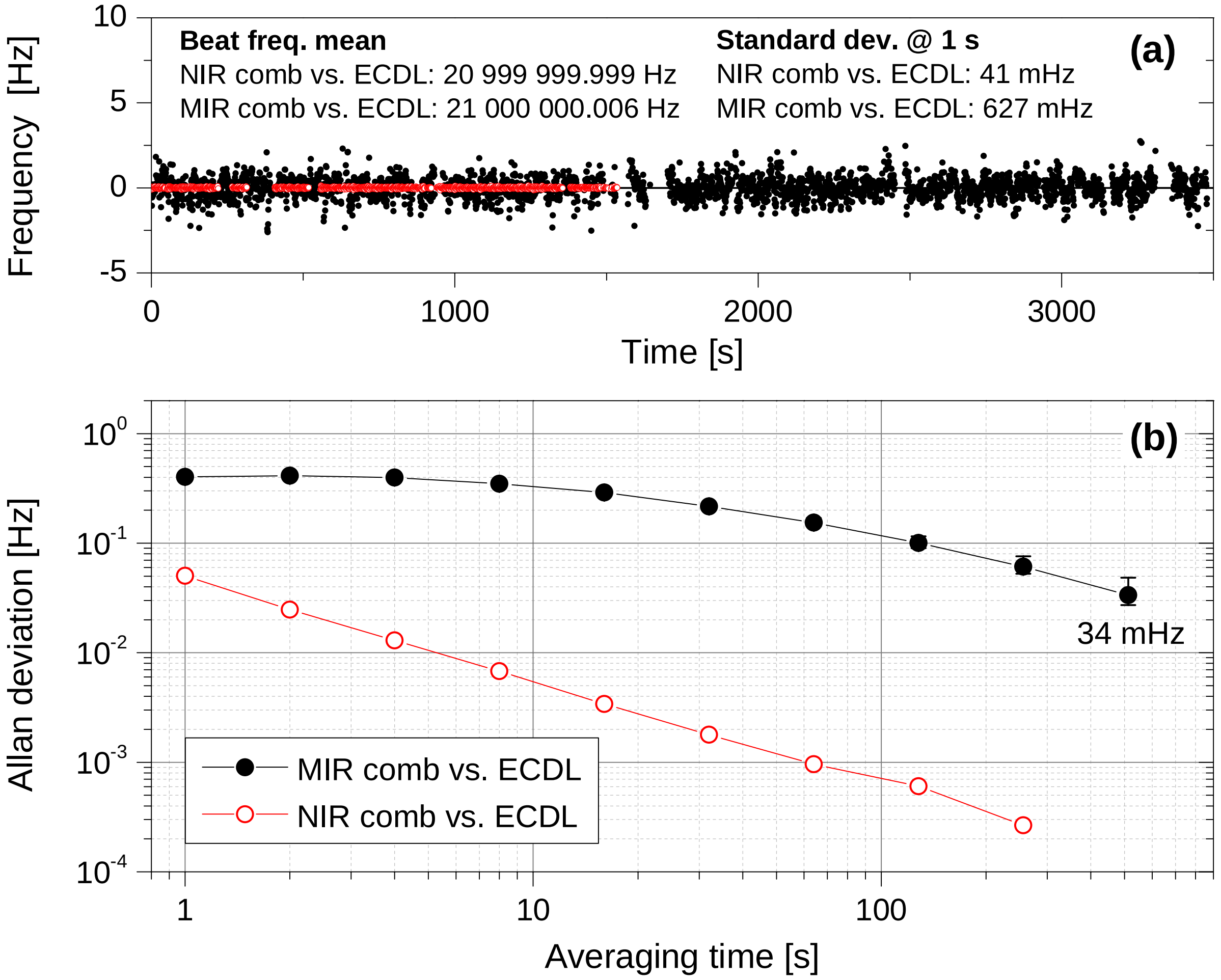

**Figure 7.** (a) Difference of the measured beat frequency from the mean value as a function of measurement time. Solid black dots: Frequency-doubled MIR comb vs. ECDL. Open red circles: NIR pump comb vs. ECDL. A counter gate time of 1 s was used in the measurements. The gaps in the data are due to occasional interruptions of phase locking between the ECDL and the pump comb. (b) The Allan deviation plots calculated from the time traces of panel (a).

## 4. Application example

To demonstrate the applicability of the fully stabilized MIR comb to precision spectroscopy, we performed comb-assisted saturation spectroscopy of methane ($CH_4$) at 3.25 µm using a mid-infrared continuous-wave OPO (CW-OPO), which was directly locked to the MIR comb. The measurement setup is depicted schematically in **Fig. 8** and the CW-OPO is described in detail in Ref. [28]. The CW-OPO is based on a singly resonant design with a 5 cm long MgO:PPLN crystal (HC Photonics), which is placed within a bow-tie resonator. In the experiments reported here, the resonator included a YAG etalon to suppress mode hops. The CW-OPO wavelength can be tuned from 2.5 to 3.5 µm or from 3.4 to 4.4 µm in a few seconds simply by tuning the wavelength of the pump laser, which is a continuous-wave titanium sapphire ring-laser (MBR-PS, Coherent). The selection between these two regions is done by poling period of the MgO:PPLN crystal of the CW-OPO.

In order to establish frequency locking between the CW-OPO and the MIR comb, their beat note was measured with a thermoelectrically cooled HgCdTe photodetector (PVI-2TE-5, VIGO), with a bandwidth up to about 30 MHz. The beat frequency was converted into a voltage signal with a frequency-to-voltage converter circuit and used as an error signal for an integrator amplifier controller. The controller tunes the CW-OPO cavity length via a PZT actuator attached to one of the cavity mirrors, thus locking the CW-OPO frequency to the comb. The linewidth of the comb-stabilized CW-OPO was measured to be about 1.5 MHz in 1-s time scale. The CW-OPO wavelength was also monitored with a wavemeter (WA-1500, EXFO, resolution ~ 40 MHz), to help with determination of the MIR comb mode number *n* and to check whether the MIR comb carrier envelope offset frequency is $f_{0,p}/2$ or $f_{0,p}/2 + f_r/2$. Sufficient accuracy in the comb offset and mode number determination was guaranteed by calibrating the wavemeter using known transitions of acetylene [29] as well as reference values for the methane transition under study. [30, 31]

Saturation spectroscopy was performed by traversing the CW-OPO beam (~15 mW) through a 50 cm gas cuvette containing 30 mTorr of methane. After the cuvette, a gold mirror reflected the beam back through the cuvette and the reflected power was measured with a second HgCdTe detector (PVM-2TE-10.6, VIGO). Owing to a fairly noisy background caused by etalon effects and instabilities in the CW-OPO power, the Lamb dip was too weak to be detected directly. Therefore, the detector signal was sent to a lock-in amplifier for phase-coherent detection. The wavelength modulation for the lock-in detection was generated by varying the MBR-PS cavity length at 4.67 kHz through the cavity tweeter mirror. The modulation depth corresponded to a few megahertz in the CW-OPO frequency. The CW-OPO wavelength was scanned over the peak of the $F_1^{(2)}$ component of the R(5) transition line of the anti-symmetric stretching vibration of $CH_4$, at about 3250.39 nm. The resulting lock-in signal of the Lamb dip is shown Fig. 8. The scanning was done by tuning the repetition rate of the NIR comb, which was used to pump the SP-OPO, in steps of 0.25 Hz. This corresponds to a change of about 100 kHz in the optical frequency of the CW-OPO that was locked to the SP-OPO comb. Without locking to the comb, the jitter of the free-running CW-OPO would be too large to reliably observe the weak Lamb dip even with lock-in detection.

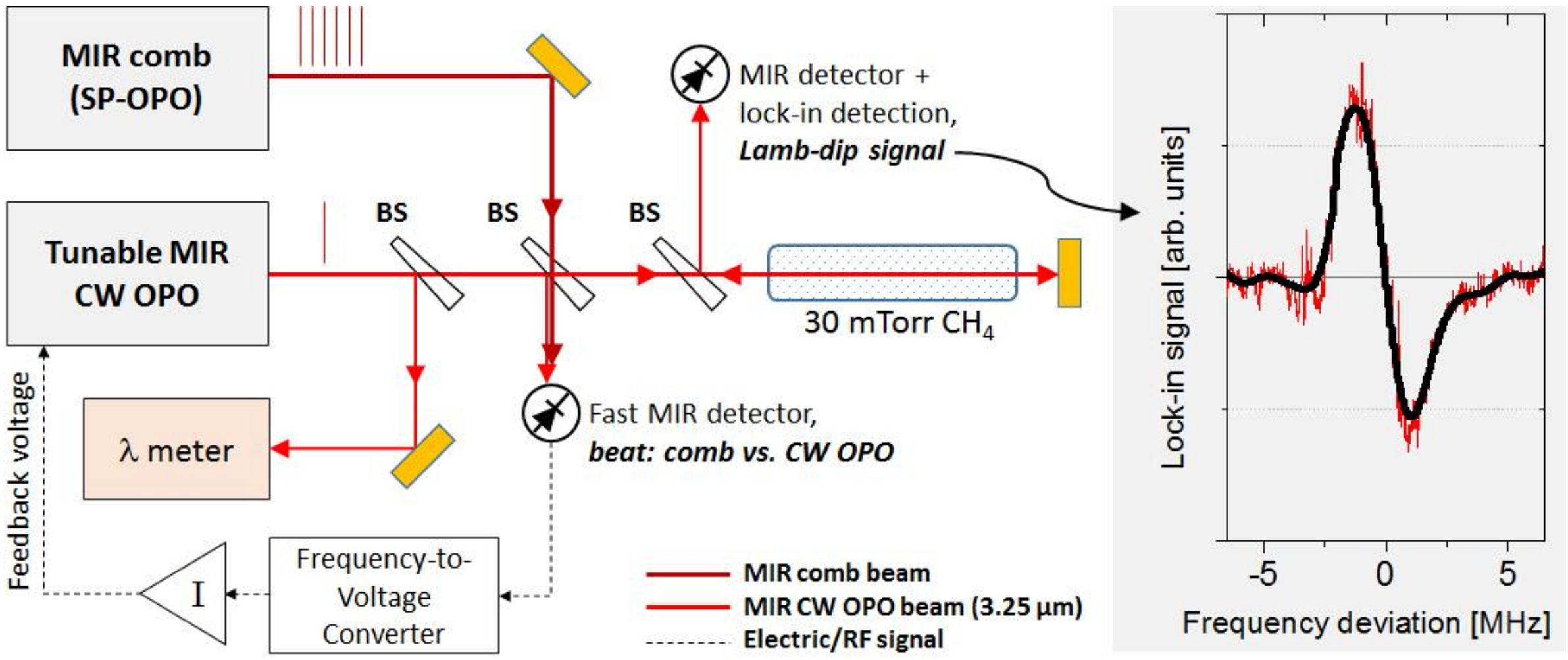

**Figure 8.** Setup of the saturation spectroscopy experiment carried out with a tunable CW OPO, which was locked to the fully stabilized MIR comb. BS = beam splitter. The inset shows a 1*f*-signal of the lock-in detected Lamb dip of methane at 3.25 µm. The measured signal is shown by the thin red trace. The thick black line is the same signal after low-pass filtering.

The full width at half maximum of the Lamb dip determined from the comb-assisted measurement is about 3.95 MHz, assuming a Lorentzian profile. The linewidth is limited mainly by the CW-OPO linewidth, but modulation broadening and residual Doppler-broadening from wavefront mismatch also contribute. The measured line center is at 92 232 636.917 MHz. This is consistent with a previously reported sub-Doppler measurement of 92 232 637.853 MHz [30] as well as with a Doppler-limited FTIR measurement of 92 232 636.94 MHz. [31] The small deviation of our result from that of Ref. [30] is likely due to the limited signal-to-noise ratio and due to systematic (technical) frequency shifts, which were not characterized for our setup.

## 5. Conclusion

In conclusion, we have demonstrated the first experimental realization and rigorous characterization of a fully stabilized mid-infrared frequency comb based on Er:fiber-laser pumped degenerate SP-OPO. The SP-OPO produces a continuous and smooth output spectrum that spans from 2.9 to 3.4 µm. We have shown that the frequencies of mid-infrared comb teeth are locked to the near-infrared pump comb with a precision of better than 20 mHz. The new MIR comb has been combined with a tunable CW OPO for comb-assisted absolute-frequency spectroscopy of methane at 3.25 µm.

**Acknowledgements** We thank Professor Lauri Halonen for support and helpful discussions during the project. The Academy of Finland and the Finnish Funding Agency for Technology and Innovation (TEKES) are acknowledged for funding. J. Karhu is grateful for postgraduate funding from the CHEMS doctoral programme of the University of Helsinki.